\documentclass[conference]{IEEEtran}
\IEEEoverridecommandlockouts
\usepackage{cite}
\usepackage{graphicx}
\usepackage{amsmath,amssymb}
\usepackage{verbatim} 
\usepackage{hyperref}

\begin{document}

\title{Development of a Quantum-Resistant File Transfer System with Blockchain Audit Trail}

\author{\IEEEauthorblockN{Ernesto Sola-Thomas}
\IEEEauthorblockA{\textit{Dept. of Electrical and Computer Engineering} \\
\textit{Clarkson University} \\
Potsdam, NY, USA \\
schumae@clarkson.edu}
\and
\IEEEauthorblockN{Masudul H Imtiaz}
\IEEEauthorblockA{\textit{Dept. of Electrical and Computer Engineering} \\
\textit{Clarkson univeristy} \\
Potsdam, NY, USA \\
mimtiaz@clarkson.edu}
}

\maketitle

\begin{abstract}
This paper presents a condensed system architecture for a file transfer system that leverages post-quantum cryptography and blockchain technology to secure data against future quantum threats. The architecture integrates NIST-standardized algorithms (CRYSTALS-Kyber for encryption and CRYSTALS-Dilithium for digital signatures) with an immutable blockchain ledger to provide an auditable, decentralized storage solution. The design is modular, comprising a Sender module for secure file encryption and signing, a central User Storage module that manages decryption, re-encryption, and blockchain logging, and a Requestor module for authenticated data retrieval. Detailed pseudocode, security considerations, and performance insights are discussed to illustrate the system's robustness, scalability, and transparency.
\end{abstract}


\section{Introduction}

The advent of quantum computing poses significant threats to traditional cryptographic methods, such as RSA (Rivest, Shamir, Adleman) and ECC (Elliptic Curve Cryptography) \cite{nist_pqc_announcement,nist_pqc_standards}. At the same time, centralized data storage models are increasingly vulnerable to breaches and misuse. The system architecture presented in this paper addresses these challenges by integrating quantum-resistant cryptographic primitives with blockchain technology.

Blockchain is a decentralized, append-only ledger maintained by a network of nodes, where each block of data is cryptographically linked to the previous one. This structure ensures transparency, immutability, and resistance to tampering—making it ideal for secure and verifiable record-keeping.

Using CRYSTALS-Kyber and CRYSTALS-Dilithium future-proofs encryption against quantum adversaries and ensures data integrity through robust digital signatures. Moreover, the immutable blockchain ledger guarantees a tamper-evident record of all file transactions, enhancing both security and regulatory compliance \cite{chainalysis_blockchain,techtarget_blockchain}.

The emergence of quantum computers represents a paradigm shift in computational capability that creates an urgent security challenge for data protection. Although large-scale quantum computers do not yet exist, adversaries may implement a ``harvest now, decrypt later" strategy, collecting currently encrypted data to decode when quantum computing becomes more accessible \cite{nist_pqc_standards}. This threat particularly affects public-key cryptographic systems like RSA and ECC, which rely on mathematical problems that quantum algorithms can solve efficiently. Shor's algorithm \cite{mamatha2024post, moody2022crypto}, for instance, can factor large numbers and solve discrete logarithm problems in minimal time, effectively breaking these widely deployed cryptographic systems. As nations and corporations invest billions in quantum computing research, the timeline for practical quantum computers capable of breaking current encryption standards continues to accelerate, creating urgency for quantum-resistant solutions.

Traditional centralized storage models compound these vulnerabilities by concentrating data under single-authority control, limiting user autonomy over access patterns and privacy. When users upload data to conventional cloud services, they surrender direct control and must trust providers to enforce access policies correctly. This concentration creates both security and compliance challenges, particularly as regulations like GDPR (General Data Protection Regulation) and CCPA (California Consumer Privacy Act) emphasize data sovereignty and transparency. Furthermore, centralized architecture presents a single point of failure that, when compromised, can lead to massive data breaches affecting millions of users. The research addresses these dual challenges by combining post-quantum cryptography with blockchain technology to create a secure, auditable, and user-centric data storage solution.

Recently, the National Institute of Standards and Technology (NIST) has led standardization efforts for post-quantum cryptographic algorithms, selecting CRYSTALS-Kyber for key encapsulation and CRYSTALS-Dilithium for digital signatures \cite{nist_pqc_announcement}. These lattice-based algorithms derive security from mathematical problems presumed difficult even for quantum computers. The architecture incorporates these NIST-standardized algorithms to ensure long-term data confidentiality and integrity against both classical and quantum adversaries. By integrating these algorithms with blockchain technology for immutable audit trails, the system offers a comprehensive security framework that addresses both present and future threats to sensitive data while enhancing regulatory compliance through transparent, verifiable records of all data transactions.

The following sections of the paper present the condensed system architecture for a quantum-resistant file transfer system with blockchain audit trails. Following this introduction, we describe the three-module system design comprising the Sender, User Storage with Blockchain, and Requestor components. We then detail implementation aspects and performance insights from the experimental evaluation. Security considerations are discussed before presenting conclusions and directions for future work. The research demonstrates that quantum-resistant security and decentralized storage principles can be practically implemented without significant performance degradation, thereby providing a viable path forward for secure data management in the post-quantum era.

\section{System Architecture}
The proposed system is divided into three primary modules: the Sender, the User Storage with Blockchain, and the Requestor. The architecture is designed to ensure secure file transmission, decentralized storage, and  auditability.

\subsection{High-Level Overview of the 3-System Design}
\begin{figure*}[htbp]
\centering
\includegraphics[width=0.9\textwidth]{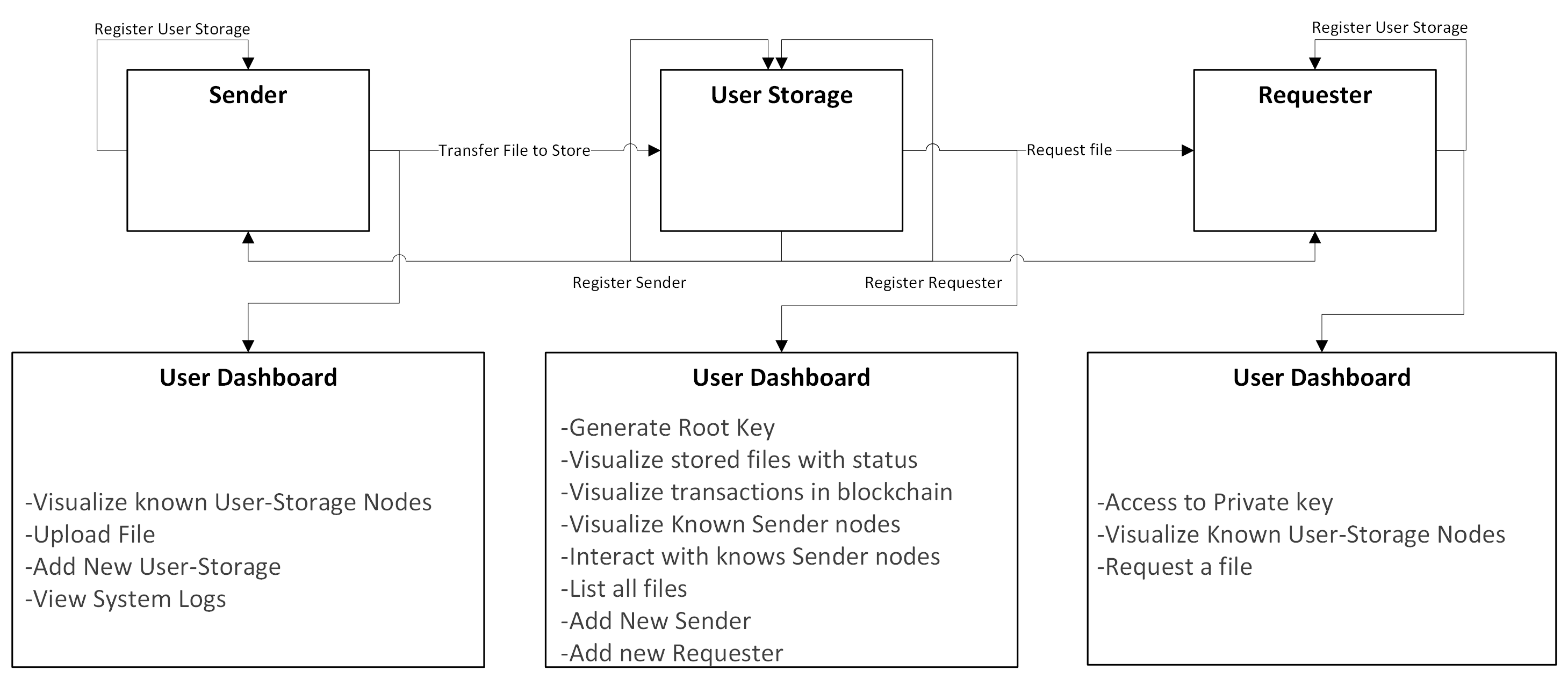}
\caption{High-level system architecture showing interactions among the Sender, User Storage with Blockchain, and Requestor modules.}
\label{fig:system_architecture}
\end{figure*}

Figure~\ref{fig:system_architecture} provides a high-level overview of the three-system design. The figure illustrates the interactions among the three key components:
\begin{itemize}
    \item \textbf{Sender Module:} Initiates secure file transfers by encrypting files with CRYSTALS-Kyber and signing them with CRYSTALS-Dilithium. Metadata such as timestamps and sender identifiers are attached to each file to support subsequent verification.
    \item \textbf{User Storage and Blockchain Module:} Acts as the central hub for processing incoming files. It decrypts files for verification, re-encrypts them using a centralized root key for secure storage, and logs each transaction on an immutable blockchain ledger. This logging captures critical details like sender ID, file name, timestamps, and transaction status, ensuring a verifiable audit trail.
    \item \textbf{Requestor Module:} Facilitates authenticated file retrieval by verifying digital signatures and re-encrypting files with the Requestor’s public key before transmission. The module also provides a user-friendly interface for managing downloads and viewing file metadata.
\end{itemize}
This modular design supports scalability and robustness by decoupling core functionalities while maintaining secure interactions across all components.

\subsection{Sender Module}
The Sender module is tasked with preparing files for secure transmission. Its core functionalities include:
\begin{itemize}
    \item \textbf{Encryption:} Files are encrypted using CRYSTALS-Kyber, a lattice-based algorithm that provides quantum resistance.
    \item \textbf{Digital Signing:} Each file is signed using CRYSTALS-Dilithium to ensure authenticity and non-repudiation.
    \item \textbf{Metadata Attachment:} Essential metadata (e.g., timestamps, sender IDs) is appended to the encrypted package to support later verification.
\end{itemize}
These processes ensure that transmitted data remains confidential and verifiable.

\subsection{User Storage and Blockchain Module}
Central to the system is the User Storage module, which functions as a secure hub for both file management and audit logging. Its key responsibilities include:
\begin{itemize}
    \item \textbf{Decryption and Re-encryption:} Files received from the Sender are decrypted and then re-encrypted using a centralized root key, ensuring uniform protection in storage.
    \item \textbf{Blockchain Logging:} Each file transaction is immutably recorded on a blockchain ledger. The ledger stores key details such as the sender ID, file name, timestamp, and status, creating a verifiable audit trail \cite{chainalysis_blockchain,techtarget_blockchain}.
    \item \textbf{Key Management:} The module manages the generation, rotation, and secure storage of quantum-resistant keys, thereby maintaining long-term system security.
\end{itemize}

Figure~\ref{fig:blockchain_logging_pseudocode} shows pseudocode outlining the blockchain logging process for sender transactions.

\begin{figure*}[htbp]
\centering
\begin{minipage}{0.9\textwidth}
\begin{verbatim}
IF HTTPS transaction received:
    Extract {encrypted_file, ciphertext, digital_signature} from Sender
    IF sender is recognized:
         Retrieve corresponding private_key
         decrypted_file = Decrypt(encrypted_file, private_key)
         VERIFY digital_signature using Dilithium
         reencrypted_file = Encrypt(decrypted_file, internal_root_key)
         Update blockchain with {sender_id, file_name, timestamp, status}
         Store file on disk as "<sender_id>_<file_name>.<ext>"
\end{verbatim}
\end{minipage}
\caption{Pseudocode outlining the blockchain logging process for sender transactions.}
\label{fig:blockchain_logging_pseudocode}
\end{figure*}

\begin{figure*}[htbp]
\centering
\includegraphics[width=0.9\textwidth]{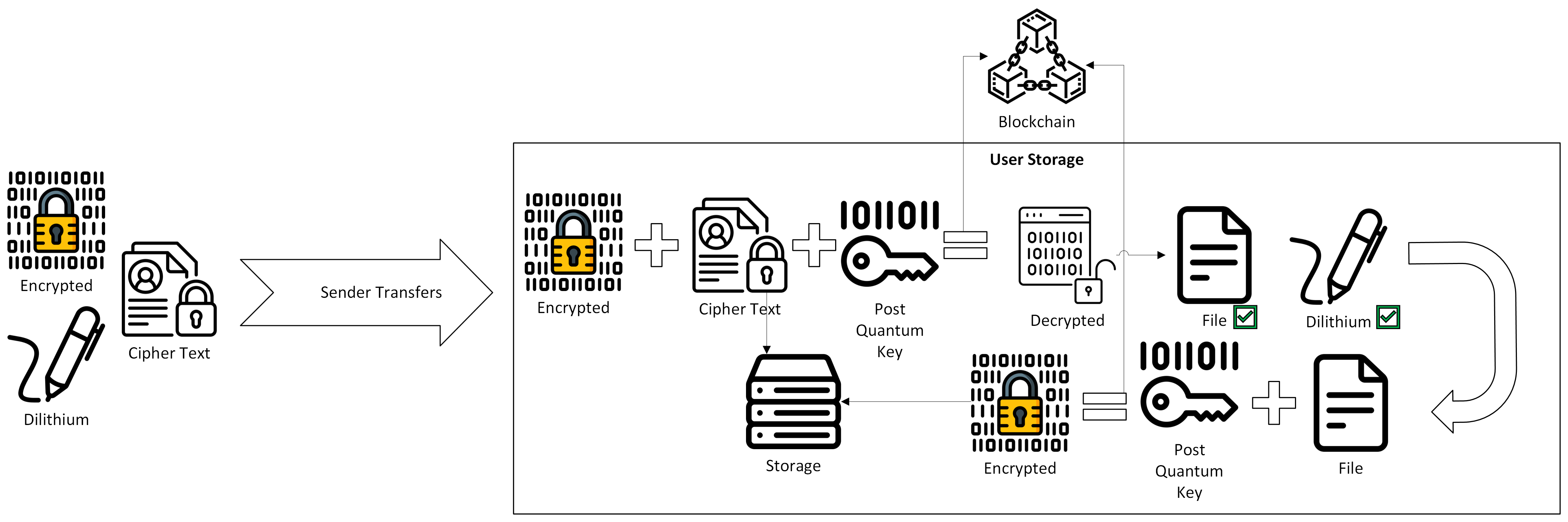}
\caption{Data flow for sender transactions processed by the User Storage module.}
\label{fig:user_storage_sender_flow}
\end{figure*}

Figure~\ref{fig:user_storage_sender_flow} depicts the detailed data flow for sender transactions processed by the User Storage module.

\subsection{Requestor Module}
The Requestor module provides secure access to stored files for authenticated users. Its operations include:
\begin{itemize}
    \item \textbf{Authentication:} Each request is authenticated using digital signatures, ensuring that only legitimate users can initiate file retrieval.
    \item \textbf{Re-encryption for Secure Transfer:} Upon verification, the User Storage module re-encrypts the requested file with the Requestor’s public key, thereby safeguarding the file during transmission.
    \item \textbf{User Interface Management:} A web-based interface allows users to view file metadata, track requests, and manage downloads.
\end{itemize}

\begin{figure*}[htbp]
\centering
\includegraphics[width=0.9\textwidth]{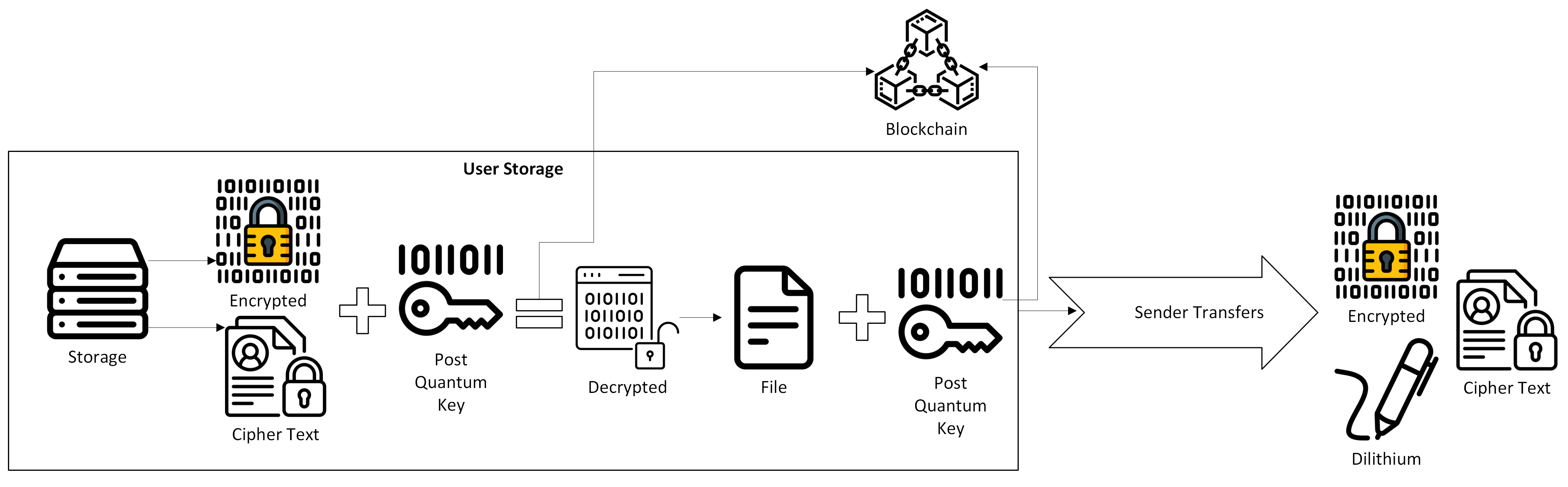}
\caption{Data flow for Requestor transactions handled by the User Storage module.}
\label{fig:user_storage_requestor_flow}
\end{figure*}

Figure~\ref{fig:user_storage_requestor_flow} illustrates the data flow for file retrieval handled by the User Storage module.

\begin{figure*}[htbp]
\centering
\includegraphics[width=0.9\textwidth]{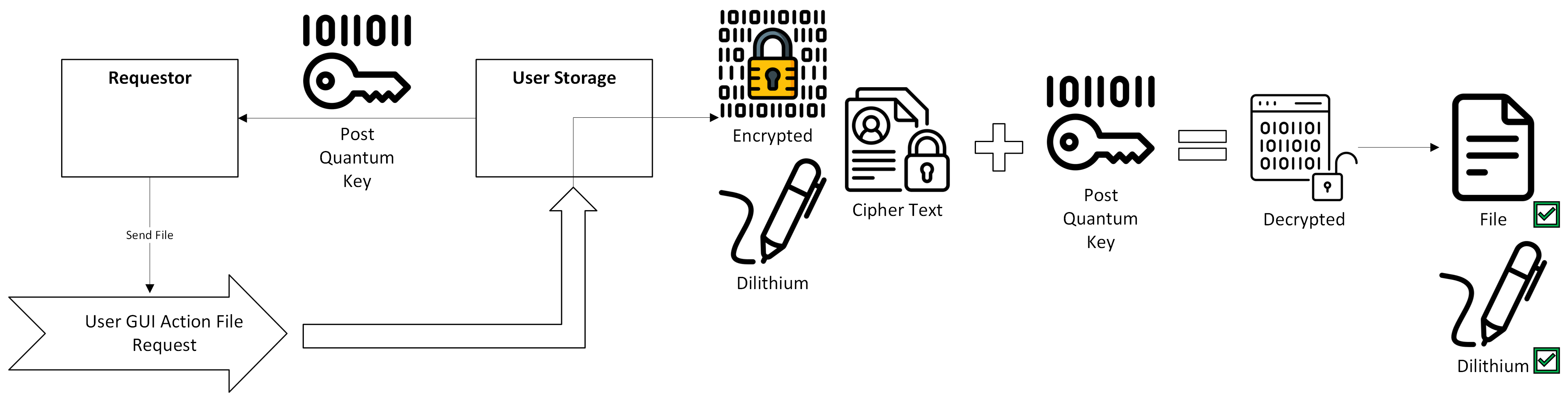}
\caption{Data flow diagram for the Requestor module.}
\label{fig:requestor_data_flow}
\end{figure*}

\section{Implementation and Performance Insights}
\begin{figure}
    \centering
    \includegraphics[width=1\linewidth]{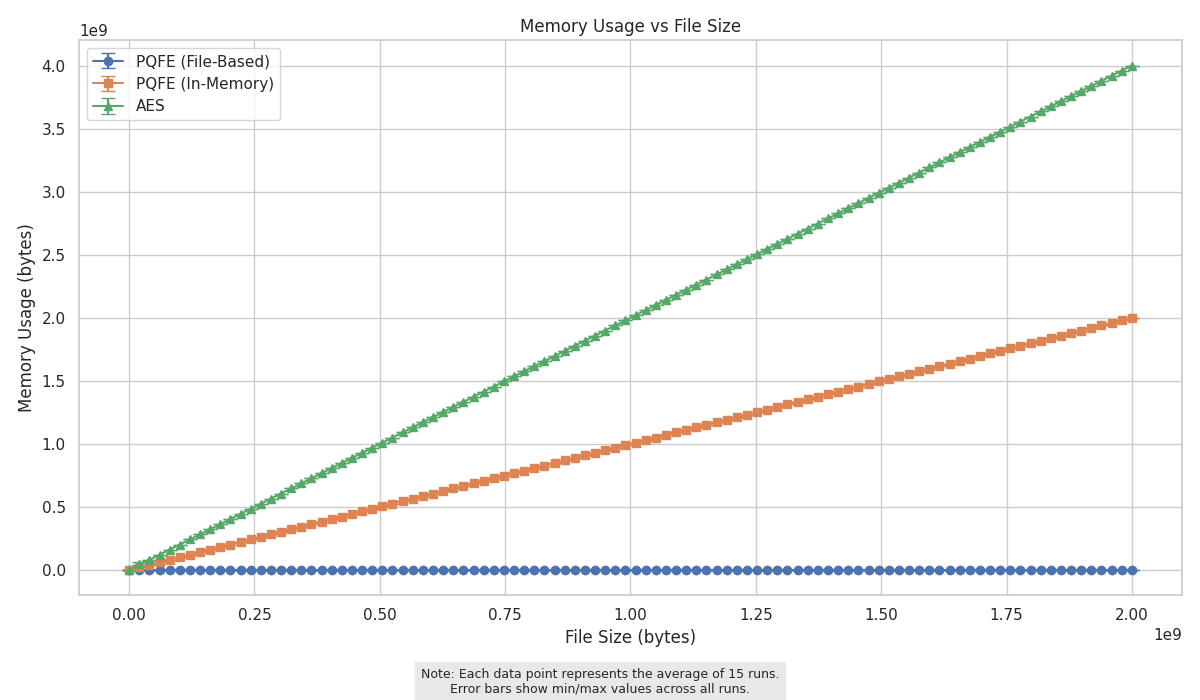}
    \caption{Peak RAM Usage vs. File Size for PQFE File-Based, PQFE In-Memory, and AES}
    \label{fig:RAM_graph}
\end{figure}

The practical implementation of the system is designed with cross-platform compatibility in mind. The Sender module is implemented as a Python web application leveraging the \texttt{liboqs} library for quantum-resistant cryptographic operations. The User Storage module is built on the Django framework with containerized deployment using Docker, ensuring consistent runtime behavior across Windows, macOS, and Linux environments.
\begin{figure}
    \centering
    \includegraphics[width=1\linewidth]{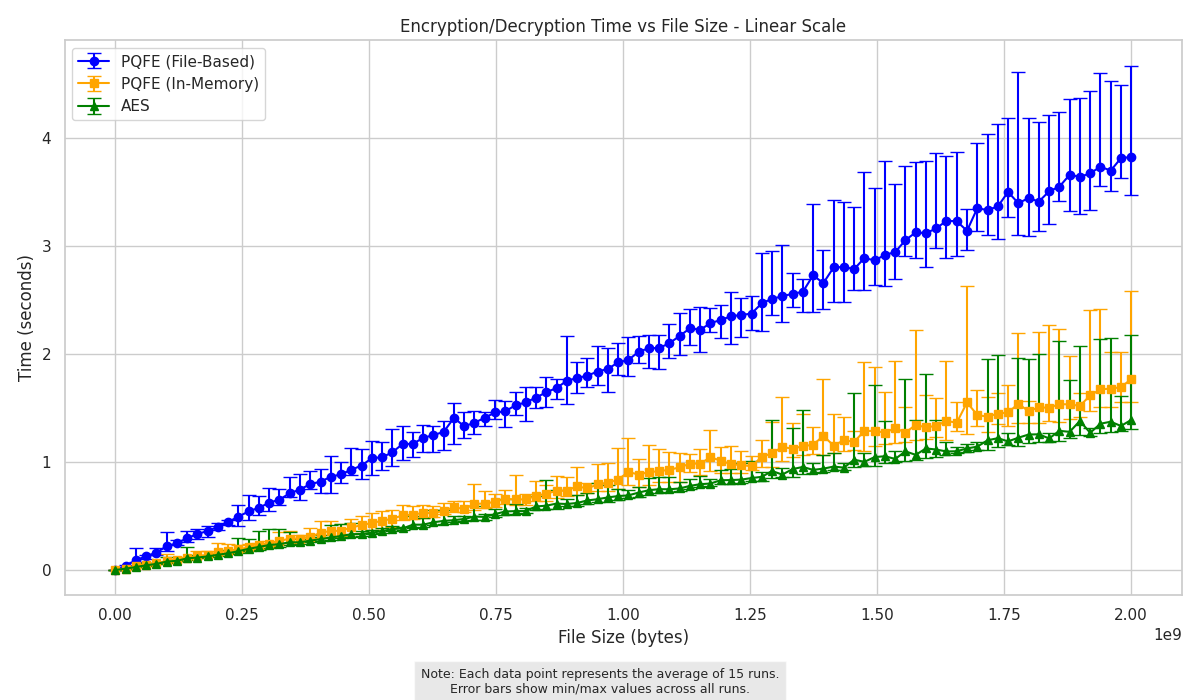}
    \caption{Time to Encrypt and Decrypt Files of Varying Sizes}
    \label{fig:Time_graph}
\end{figure}
Experimental evaluations were performed on a single-core x86 system with 16 GB of RAM. Files ranging from 1 KB to 2 GB were processed in both PQC In-Memory and file-based modes. Benchmarking tests revealed that while the file-based approach incurs a modest latency increase (e.g., 3.8 seconds for a 2GB file) ~\ref{fig:Time_graph}, the In-Memory approach achieves nearly identical performance to traditional AES encryption (1.3 seconds versus 1.2 seconds for 2GB files) \cite{demir2023performance,bavdekar2022post}. These findings highlight the feasibility of deploying post-quantum algorithms in practical, low-resource environments, ~\ref{fig:RAM_graph} highlights RAM utilization of PQFE and AES. PQFE FIle-Based reading from disk achieving constant to zero RAM utilization with compute time increase see in ~\ref{fig:Time_graph} and PQFE in-Memory processing encryption/decryption in RAM beating the standard AES algorithm in terms of peak RAM utilization.
\section{Security Considerations}
The integration of CRYSTALS-Kyber and CRYSTALS-Dilithium enhances data confidentiality and integrity against both classical and quantum adversaries \cite{nist_pqc_announcement,nist_pqc_standards}. Additionally, blockchain logging serves as an immutable audit trail that records all file transactions, facilitating rapid detection of unauthorized modifications and ensuring compliance with regulatory requirements. The system's modular design further simplifies the implementation of advanced key management techniques, including key rotation and dynamic protocol negotiation.

\section{Discussion and Future Work}
The presented architecture demonstrates a robust approach to secure file transfer in a post-quantum era. Key benefits include:
\begin{itemize}
    \item \textbf{Enhanced Security:} Quantum-resistant cryptographic algorithms ensure long-term data confidentiality.
    \item \textbf{Auditability:} Blockchain-based logging provides an immutable and transparent record of all transactions.
    \item \textbf{Decentralized Control:} The system empowers users by reducing reliance on centralized storage solutions.
\end{itemize}

The research findings contradict initial concerns over lattice-based cryptography’s overhead. PQFE In-Memory encryption runs nearly as fast as AES (1.3 vs. 1.2 seconds for 2GB files), while File-Based encryption’s 3.8 seconds remains acceptable \ref{fig:Time_graph} ~\ref{fig:RAM_graph}. These results confirm that lattice-based schemes are practical, offering both quantum resistance and blockchain audit trails. The modular design also streamlines maintenance and enables targeted optimizations.

Future work will focus on optimizing multi-node blockchain deployments, refining key management protocols, and integrating hardware acceleration techniques (e.g., FPGA or ASIC) to further reduce latency and improve throughput. Additionally, efforts will be made to enhance the graphical user interfaces and extend the system's scalability to support higher transaction volumes in real-world applications. Specific research directions include developing a fully decentralized ledger with dynamic resource allocation, designing custom ASIC implementations for PQC operations, and establishing automated cryptographic agility systems to allow seamless algorithm updates as cryptographic standards evolve. While the current implementation serves as a proof of concept, these enhancements will transform it into a production-ready solution capable of meeting enterprise-scale security and performance requirements in the emerging post-quantum landscape. The first application area will be tested by integrating this technology into our previous health and biometric technologies ~\cite{alam2024nir,sarker2024face,sola2023fpga,mooney2023sensory,imtiaz2017wearable,sarker2023wheelchair,sola2021smartwheelchair,sola2021stereo}. 

\section{Conclusion}
In conclusion, this paper has presented a condensed yet comprehensive system architecture for a quantum-resistant file transfer system with blockchain audit trails. By combining advanced cryptographic techniques with decentralized storage principles, the proposed design addresses both current and emerging security challenges. The experimental evaluations affirm that the computational overhead introduced by post-quantum algorithms is manageable, paving the way for future research into hardware-accelerated solutions and multi-node blockchain scalability.

\section{Attribution}
\subsection{Icons and Graphics}
The icons used in the diagrams were downloaded from Flaticon:
\begin{itemize}
    \item \url{https://support.flaticon.com/}
\end{itemize}

\vspace{12pt}

\end{document}